\documentclass{article}
\usepackage{ijcai19}
\usepackage[english]{babel}

\usepackage{times}
\usepackage{soul}
\usepackage{url}
\usepackage[utf8]{inputenc}
\usepackage[small]{caption}
\usepackage{graphicx}
\usepackage{amsmath}
\usepackage{booktabs}
\usepackage{graphicx}
\title{Automatic Grading of Knee Osteoarthritis on the Kellgren-Lawrence Scale from Radiographs Using Convolutional Neural Networks}

\author{
Sudeep Kondal\and
Viraj Kulkarni\and
Ashrika Gaikwad\and
Amit Kharat\and
Aniruddha Pant
\affiliations
DeepTek Inc\\
\emails
}

\begin{document}

\maketitle

\begin{abstract}
The severity of knee osteoarthritis is graded using the 5-point Kellgren-Lawrence (KL) scale where healthy knees are assigned grade 0, and the subsequent grades 1-4 represent increasing severity of the affliction. Although several methods have been proposed in recent years to develop models that can automatically predict the KL grade from a given radiograph, most models have been developed and evaluated on datasets not sourced from India. These models fail to perform well on the radiographs of Indian patients. In this paper, we propose a novel method using convolutional neural networks to automatically grade knee radiographs on the KL scale. Our method works in two connected stages: in the first stage, an object detection model segments individual knees from the rest of the image; in the second stage, a regression model automatically grades each knee separately on the KL scale. We train our model using the publicly available Osteoarthritis Initiative (OAI) dataset and demonstrate that fine-tuning the model before evaluating it on a dataset from a private hospital significantly improves the mean absolute error from 1.09 (95\% CI: 1.03-1.15) to 0.28 (95\% CI: 0.25-0.32). Additionally, we compare classification and regression models built for the same task and demonstrate that regression outperforms classification.
\end{abstract}

\section{Introduction}
Knee osteoarthritis is a painful joint disorder that obstructs the natural movement of the knee. Its major indications include joint space narrowing, osteophytes formation, and sclerosis \cite{kellgren1957radiological}. It is most commonly observed in people who are above the age of 45, suffer from obesity, or lead a sedentary lifestyle. Since no known cure exists for reversing knee osteoarthritis, early diagnosis is crucially important for a patient to prevent its further progression by making behavioral and lifestyle changes \cite{10.3389/fmech.2019.00057}. Despite the availability of advanced medical imaging techniques such as Magnetic Resonance Imaging (MRI) and Computed Tomography (CT), the knee radiograph remains the most widely used modality for the diagnosis of osteoarthritis. The severity of the disease is graded on the Kellgren-Lawrence (KL) scale \cite{kellgren1957radiological} where grade 0 suggests no radiographic features of OA are present; grade 1 suggests doubtful joint space narrowing (JSN) and possible osteophytic lipping; grade 2 suggests definite osteophytes and possible JSN on anteroposterior weight-bearing radiograph; grade 3 suggests multiple osteophytes, definite JSN, sclerosis, possible bony deformity; and grade 4 suggests large osteophytes, marked JSN, severe sclerosis and definite bony deformity \cite{kohn2016classifications}.

Although it is widely used, the KL grading system suffers from reader subjectivity in evaluating radiographs \cite{gossec2008comparative}\cite{sheehy2015validity}\cite{culvenor2015defining}. Computer-aided diagnosis can mitigate this subjectivity by providing an automated assessment to assist radiologists in making the final decision. In this study, we propose a method for building a convolutional neural network (CNN) system to automatically grade a knee radiograph on the KL scale. We first develop a preprocessing network that takes as input a knee radiograph and outputs two images: one image containing only the left knee, and another image containing only the right knee. The second network takes as input a single knee image and outputs the predicted KL grade. For this second network, we consider two options - classification and regression - and observe that regression performs better than classification.

Most prior work on automatically diagnosing knee osteoarthritis is based on datasets which have been collected from patients not based in India, particularly the Osteoarthritis Initiative (OAI) Dataset\cite{antony2016quantifying}\cite{antony2017automatic}\cite{tiulpin2018automatic} containing 4447 images and the Multicenter Osteoarthritis Study (MOST) Dataset \cite{antony2017automatic}\cite{tiulpin2018automatic}\cite{tiulpin2017novel} with 3026 images. We demonstrate that the model developed using OAI Dataset did not perform well on a dataset sourced from a private hospital in India (hereafter referred to as \textit{Target Dataset}). However, when the model was fine-tuned on Target Dataset, the performance improved significantly.

\section{Related Work}
Machine learning has been widely used to detect and grade knee osteoarthritis. Kotti et al. \cite{kotti2017detecting} built a regression model from data points such as mean value, push-off time, and slope of vertical, anterior–posterior, and medio-lateral ground reaction forces measured using force plates with piezoelectric 3-component force sensors. Bandyopadhyay et al. \cite{Bandyopadhyay2016Detection} argued that knee osteoarthritis should be diagnosed using features of the cartilage; they segmented the cartilage from the knee x-ray image, extracted features like statistical moments, texture, shape, etc., and used them to train a random forest \cite{breiman2001random} classifier to predict knee osteoarthritis severity. Brahim et al. \cite{brahim2019decision} used images from OAI Dataset to predict knee osteoarthritis severity using image preprocessing techniques like circular Fourier filter and multivariate linear regression and extracted important features using independent component analysis; these extracted features were further used to train naive Bayes \cite{anderson1992explorations} and random forest classifiers. Orlov et al. proposed Wndchrm \cite{orlov2008wnd}, an image classifier that uses hand-crafted features based on polynomial decomposition, contrast, pixel statistics, and textures in addition to features obtained by performing image transformations \cite{shamir2008source}\cite{shamir2009early}\cite{orlov2008wnd}.

Manually designing and selecting relevant features, however, require a significant degree of domain knowledge \cite{lee2010unsupervised}, while image processing techniques such as those in \cite{Bandyopadhyay2016Detection} are not robust and are prone to noise. Convolutional neural networks (CNN) \cite{lecun1999object}, on the other hand, learn relevant features automatically from the training images without human intervention, and they have achieved state-of-art performance on many image classification tasks in recent times \cite{sultana2018advancements}. Antony et al. \cite{antony2016quantifying}\cite{antony2017automatic} investigated the use of well-known CNNs such as the VGG 16-layer net \cite{simonyan2014very}, VGG-M-128 \cite{chatfield2014return}, and BVLC reference CaffeNet \cite{jia2014caffe} to classify knee radiographs and demonstrated that they performed better than the prevailing state-of-art methods that used hand-crafted features. Tiulpin et al. \cite{tiulpin2018automatic} used a deep Siamese convolutional neural network \cite{chopra2005learning} to classify knee radiographs into KL grades and presented attention maps highlighting the radiological features affecting the network's decisions.

Machine learning methods, in general, and neural networks, in particular, are known to face difficulties in generalizing to datasets other than the ones used to train them. We show in this study that the model trained on OAI Dataset performed poorly when used to evaluate images from Target Dataset. We further show that the performance increased dramatically when the model was subsequently fine-tuned on Target Dataset. Two approaches have been considered by prior research work: the classification approach where the different KL grades 0-4 are treated as discrete, unordered classes; and the regression approach where the KL grade is treated as a continuous response variable that takes a value in the range [0, 4]. We contrast the performance of both approaches and conclude that regression performs better than classification.

\section{Data and Method}

In this section we describe the datasets, the methods used to build the models, and the results.

\subsection{Data}
We obtained a set of 4447 knee radiographs in DICOM format along with the KL grades for both left and right knees from the Osteoarthritis Initiative (OAI). The dataset contained images of males and females between 45 and 79 years of age. From a well-known private hospital in India, we obtained Target Dataset consisting of 1043 knee radiographs. Both datasets contained images in the anterior-posterior view. Table 1 shows the distribution of samples in different KL grades for the two datasets.

\begin{table}[hbt!]
\begin{center}
\begin{tabular}{ | c | c c c c c c | }
\hline
KL Grade & 0 & 1 & 2 & 3 & 4 & Total\\ 
\hline
OAI Dataset & 3493 & 2319 & 1595 & 1177 & 310 & 8894\\ 
Target Dataset & 335 & 150 & 199 & 194 & 297 &  1175\\ 
\hline
\end{tabular}
\caption{Distribution of images into KL grades}
\end{center}
\end{table}

\subsection{Segmenting the Knee Joints from the Radiographs}
We randomly selected 1000 images from OAI Dataset. Using the VGG Image Annotator (VIA) tool \cite{dutta2019vgg}, we drew bounding boxes around each knee joint in these images. These 1000 images containing 2000 annotated knee joints were used to train a Mask R-CNN model \cite{he2017mask} that could segment the two knee joints from a given radiograph. In addition to segmenting the knees, the model was also trained to differentiate between the left and the right knee to aid the subsequent process of report generation. The Mask R-CNN algorithm works in two steps. In the first step, it generates region proposals through a region proposal network (RPN) \cite{ren2015faster}. In the second step, it predicts the class-label for each region and regresses the coordinates of the bounding box that encloses the knee joint. The model was tested on 100 randomly selected images from OAI Dataset. The mean squared error (MSE) for bounding box regression was 0.0507, the DICE score for segmenting the knees was 0.93, and the accuracy for classifying the segmented knee into left and right was 99\%. Using this model, we segmented the knee joints from the rest of the images in OAI dataset (see figure 1 and figure 2).

\begin{figure}
\includegraphics[width=\linewidth]{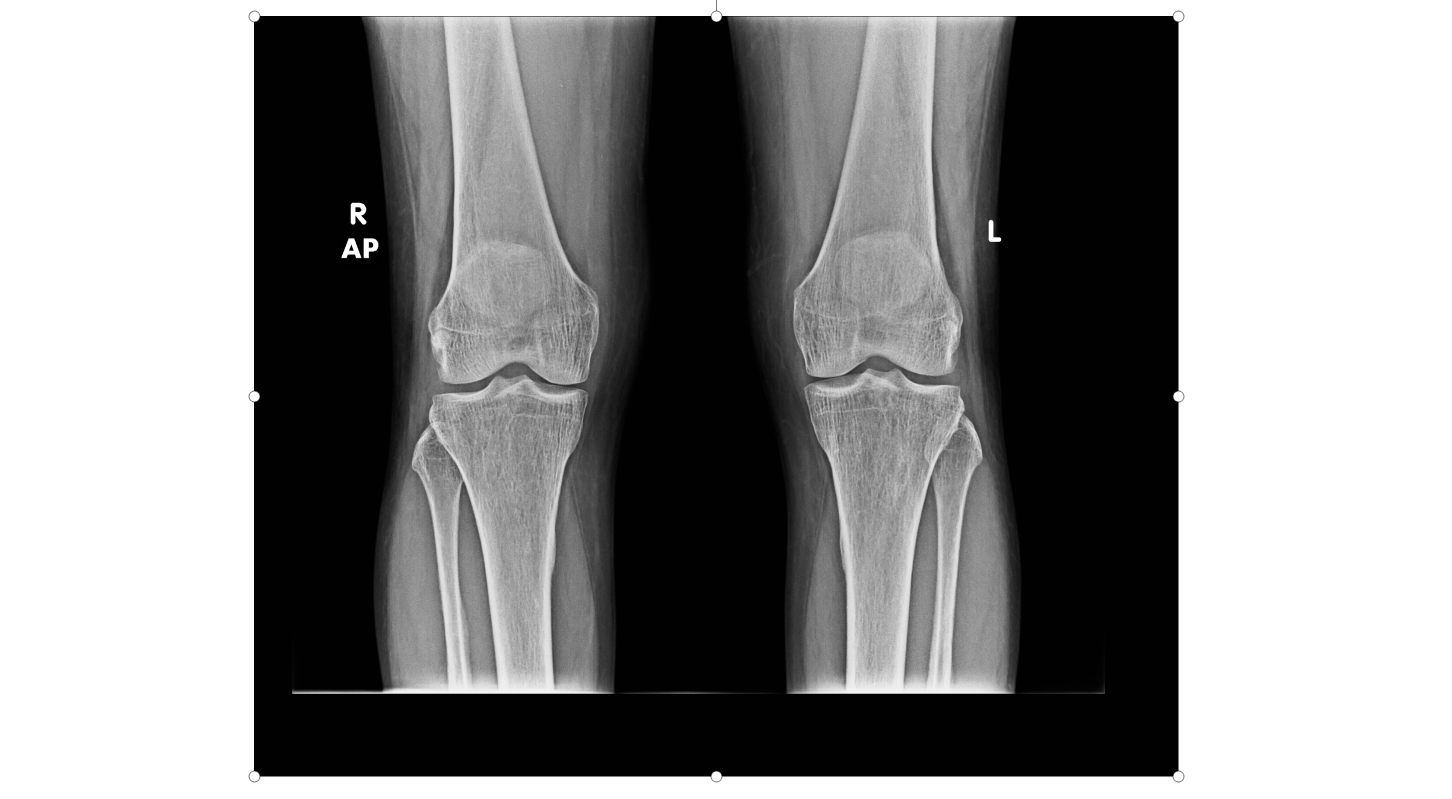}
\caption{Knee radiograph sample provided as input to knee segmentation model}
\end{figure}

\begin{figure}
\includegraphics[width=\linewidth]{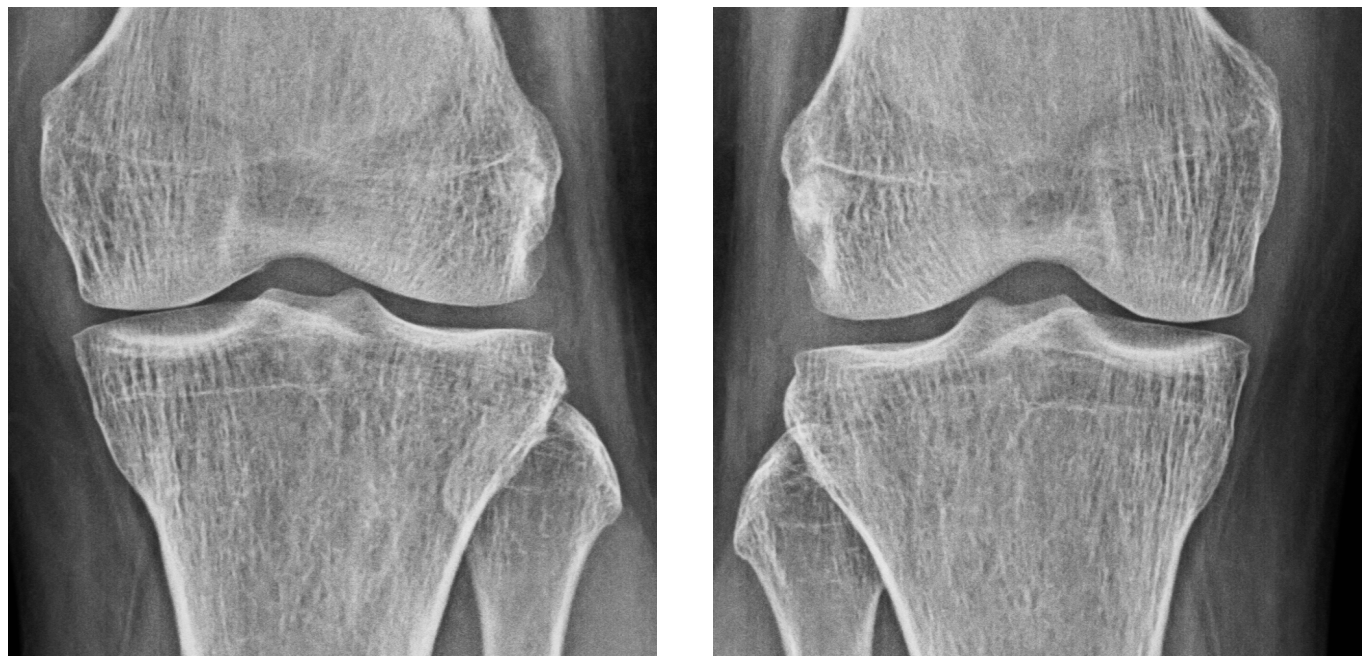}
\caption{Left and right knees as segmented from image in figure 1}
\end{figure}

\subsection{Predicting KL Grade from Segmented Knee Images}
The segmented knee images were normalized using min-max normalization to ensure the pixel intensity values were bounded in the range [0,255]. OAI Dataset was divided into training, validation, and testing subsets in the ratios 70:10:20. Grading on the KL scale is based on joint space narrowing, and the differences between two consecutive grades are very subtle in appearance. In a conventional convolutional neural network (CNN), each layer is connected only to its adjacent layer. In DenseNet \cite{huang2017densely}, however, every layer is connected to every other layer; this allows the network to learn better features from lesser data. We used the DenseNet-121 architecture to train a classification model that classified each input knee image into one class from the set of discrete, unordered classes: \{0, 1, 2, 3, 4\}. Cross-entropy was used as the loss function for training the model. We obtained a mean class-wise precision of 0.55 and a mean class-wise recall of 0.57 (see table 2 for detailed results).

\begin{table}[hbt!]
\begin{center}
\begin{tabular}{ |c | c | c | c | c | }
\hline
KL Grade & Precision & Recall & F1-Score & Kappa\\ 
\hline
0 & 0.67 & 0.88 & 0.76 & 0.57\\
1 & 0.57 & 0.01 & 0.02 & 0.02\\
2 & 0.63 & 0.47 & 0.54 & 0.41\\
3 & 0.68 & 0.54 & 0.60 & 0.55\\
4 & 0.18 & 0.95 & 0.31 & 0.27\\
Average & 0.55 & 0.57 & 0.45 & 0.36\\
\hline
\end{tabular}
\caption{Performance of classification model on OAI Dataset when evaluated on OAI Dataset}
\end{center}
\end{table}

The KL-grades present a natural ordering amongst them. To take advantage of this ordering, we replaced the output layer of the classification network with two fully connected layers. The penultimate layer had 128 nodes with ReLU activation, while the final output layer had a single node with linear activation. This model was trained using mean squared error as the loss function to predict the KL grade as a real number. The model output was rounded off to the closest integer in the range [0, 4] to obtain the final predicted grade. The mean class-wise precision was 0.84 and the mean class-wise recall was 0.82 (see table 3 for detailed results).

\begin{table}[hbt!]
\begin{center}
\begin{tabular}{ | c | c | c | c | c | }
\hline
KL Grade & Precision & Recall & F1-Score & Kappa\\ 
\hline
0 & 0.93 & 0.91 & 0.92 & 0.86\\
1 & 0.71 & 0.74 & 0.73 & 0.66\\
2 & 0.85 & 0.87 & 0.86 & 0.80\\
3 & 0.85 & 0.85 & 0.85 & 0.83\\
4 & 0.84 & 0.76 & 0.80 & 0.79\\
Average & 0.84 & 0.82 & 0.83 & 0.79\\
\hline
\end{tabular}
\caption{Performance of regression model on OAI Dataset when evaluated on OAI Dataset}
\end{center}
\end{table}

\subsection{Fine-tuning Models for Target Dataset}
When evaluated on Target Dataset, the regression model trained on OAI Dataset yielded a mean class-wise precision of 0.46 and a mean class-wise recall of 0.33 (see table 4 for detailed results). Although analyzing the reasons for this drop in performance is beyond the scope of the paper, we surmise that the images from the two datasets differed owing to reasons including, but not limited to, different image capture settings, different imaging equipment, and perhaps different patient population characteristics. The difference between OAI Dataset and Target Dataset is visually represented in figure 3.

\begin{figure}
\includegraphics[width=\linewidth]{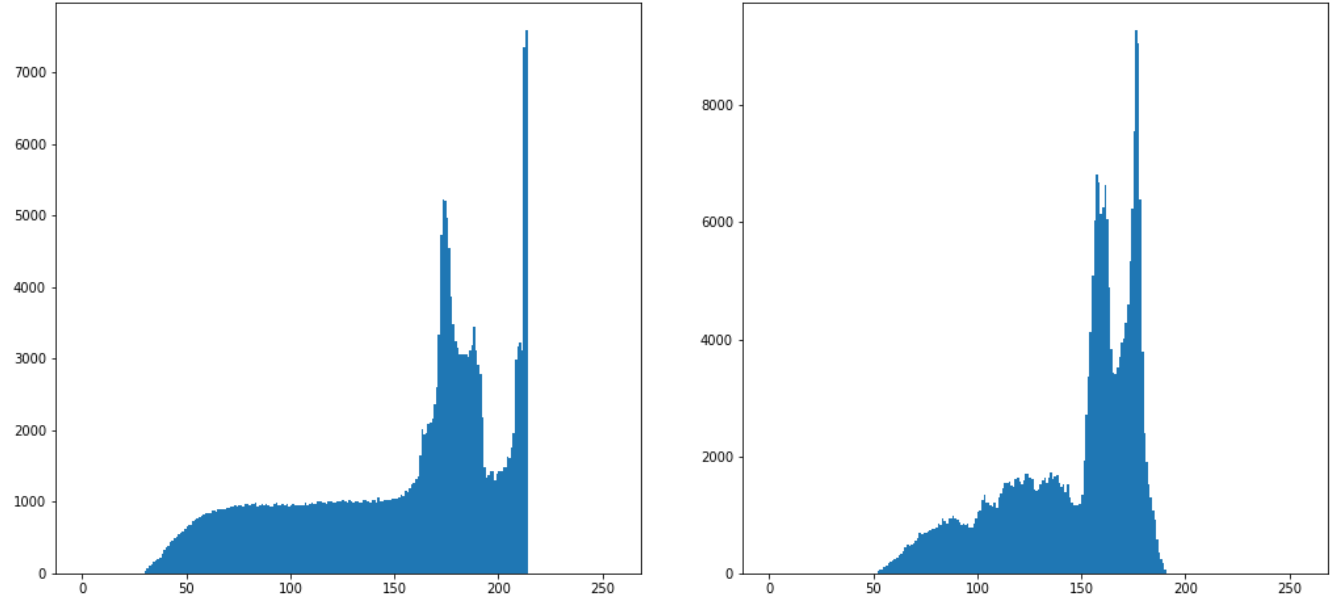}
\caption{(Left) Histogram of pixel intensities of images from OAI Dataset; (Right) Histogram of pixel intensities of images from Target Dataset}
\end{figure}

\begin{table}[hbt!]
\begin{center}
\begin{tabular}{ | c | c | c | c | c | }
\hline
KL Grade & Precision & Recall & F1-Score & Kappa\\ 
\hline
0 & 1 & 0.02 & 0.04 & 0.03\\
1 & 0.17 & 0.09 & 0.12 & 0.03\\
2 & 0.22 & 0.43 & 0.29 & 0.09\\
3 & 0.22 & 0.53 & 0.31 & 0.10\\
4 & 0.70 & 0.57 & 0.63 & 0.52\\
Average & 0.46 & 0.33 & 0.28 & 0.16\\
\hline
\end{tabular}
\caption{Performance of regression model trained on OAI Dataset when evaluated on Target Dataset}
\end{center}
\end{table}

\begin{table}[hbt!]
\begin{center}
\begin{tabular}{ |c | c | c | c | c | }
\hline
KL Grade & Precision & Recall & F1-Score & Kappa\\ 
\hline
0 & 0.90 & 0.79 & 0.84 & 0.79\\
1 & 0.52 & 0.70 & 0.59 & 0.52\\
2 & 0.63 & 0.68 & 0.65 & 0.58\\
3 & 0.67 & 0.64 & 0.65 & 0.59\\
4 & 0.91 & 0.86 & 0.88 & 0.85\\
Average & 0.73 & 0.73 & 0.73 & 0.66\\
\hline
\end{tabular}
\caption{Performance of regression model trained on OAI Dataset and fine-tuned on OAI Dataset when evaluated on Target Dataset}
\end{center}
\end{table}

We computed the mean absolute error (MAE) between the predicted KL grade and the actual KL grade. The MAE for the model trained on OAI Dataset when evaluated on Target Dataset was 1.09 (95\% CI: 1.03-1.15). The MAE for the fine-tuned model when evaluated on Target Dataset was 0.28 (95\% CI: 0.25-0.32).

\section{Discussion}
Our experiments uncovered differences in image characteristics between OAI Dataset and Target Dataset. The model trained on OAI Dataset did not perform well on Target Dataset. However, when we fine-tuned the model on Target Dataset, we observed a significant improvement in performance.

Since the Kellgren-Lawrence system presupposes an implicit ordering between its grades, we argue that the grade should be treated as an ordinal variable rather than a nominal variable. Training a classification model using cross-entropy as the loss function imposes an equal penalty regardless of whether the misclassification is between neighbouring grades or far-away grades. However, defining this as a regression problem and using mean squared error as the loss function imposes lower penalties on misclassifications between neighboring grades and progressively higher penalties as the distance between the actual and predicted grades increases. When evaluating the fine-tuned model on Target Dataset, 68\% of all misclassifications were amongst neighbouring grades for the classification model, while 87\% of all misclassifications were amongst neighbouring grades for the regression model.

Since different stages of knee osteoarthritis require different interventions, it is important to identify the grade of osteoarthritis affecting a patient. By automatically and accurately assigning KL grades to knee radiographs, the proposed method can help (1) mitigate the effects of human subjectivity in assessing radiographs; (2) reduce radiologist work burden; (3) improve reporting times.

% \printbibliography 
\bibliographystyle{ieeetr}
\bibliography{bibliography}

\begin{thebibliography}{10}

\bibitem{kellgren1957radiological}
J.~Kellgren and J.~Lawrence, ``Radiological assessment of osteo-arthrosis,''
  {\em Annals of the rheumatic diseases}, vol.~16, no.~4, p.~494, 1957.

\bibitem{10.3389/fmech.2019.00057}
X.~Ji and H.~Zhang, ``Current strategies for the treatment of early stage
  osteoarthritis,'' {\em Frontiers in Mechanical Engineering}, vol.~5, p.~57,
  2019.

\bibitem{kohn2016classifications}
M.~D. Kohn, A.~A. Sassoon, and N.~D. Fernando, ``Classifications in brief:
  Kellgren-lawrence classification of osteoarthritis,'' 2016.

\bibitem{gossec2008comparative}
L.~Gossec, J.~Jordan, S.~Mazzuca, M.-A. Lam, M.~Suarez-Almazor, J.~Renner,
  M.~Lopez-Olivo, G.~Hawker, M.~Dougados, J.~Maillefert, {\em et~al.},
  ``Comparative evaluation of three semi-quantitative radiographic grading
  techniques for knee osteoarthritis in terms of validity and reproducibility
  in 1759 x-rays: report of the oarsi--omeract task force,'' {\em
  Osteoarthritis and cartilage}, vol.~16, no.~7, pp.~742--748, 2008.

\bibitem{sheehy2015validity}
L.~Sheehy, E.~Culham, L.~McLean, J.~Niu, J.~Lynch, N.~A. Segal, J.~A. Singh,
  M.~Nevitt, and T.~D.~V. Cooke, ``Validity and sensitivity to change of three
  scales for the radiographic assessment of knee osteoarthritis using images
  from the multicenter osteoarthritis study (most),'' {\em Osteoarthritis and
  cartilage}, vol.~23, no.~9, pp.~1491--1498, 2015.

\bibitem{culvenor2015defining}
A.~G. Culvenor, C.~N. Engen, B.~E. {\O}iestad, L.~Engebretsen, and M.~A.
  Risberg, ``Defining the presence of radiographic knee osteoarthritis: a
  comparison between the kellgren and lawrence system and oarsi atlas
  criteria,'' {\em Knee Surgery, Sports Traumatology, Arthroscopy}, vol.~23,
  no.~12, pp.~3532--3539, 2015.

\bibitem{antony2016quantifying}
J.~Antony, K.~McGuinness, N.~E. O'Connor, and K.~Moran, ``Quantifying
  radiographic knee osteoarthritis severity using deep convolutional neural
  networks,'' in {\em 2016 23rd International Conference on Pattern Recognition
  (ICPR)}, pp.~1195--1200, IEEE, 2016.

\bibitem{antony2017automatic}
J.~Antony, K.~McGuinness, K.~Moran, and N.~E. O’Connor, ``Automatic detection
  of knee joints and quantification of knee osteoarthritis severity using
  convolutional neural networks,'' in {\em International conference on machine
  learning and data mining in pattern recognition}, pp.~376--390, Springer,
  2017.

\bibitem{tiulpin2018automatic}
A.~Tiulpin, J.~Thevenot, E.~Rahtu, P.~Lehenkari, and S.~Saarakkala, ``Automatic
  knee osteoarthritis diagnosis from plain radiographs: A deep learning-based
  approach,'' {\em Scientific reports}, vol.~8, no.~1, pp.~1--10, 2018.

\bibitem{tiulpin2017novel}
A.~Tiulpin, J.~Thevenot, E.~Rahtu, and S.~Saarakkala, ``A novel method for
  automatic localization of joint area on knee plain radiographs,'' in {\em
  Scandinavian Conference on Image Analysis}, pp.~290--301, Springer, 2017.

\bibitem{kotti2017detecting}
M.~Kotti, L.~D. Duffell, A.~A. Faisal, and A.~H. McGregor, ``Detecting knee
  osteoarthritis and its discriminating parameters using random forests,'' {\em
  Medical engineering \& physics}, vol.~43, pp.~19--29, 2017.

\bibitem{Bandyopadhyay2016Detection}
S.~K. Bandyopadhyay and P.~Sharma, ``Detection of osteoarthritis using knee
  x-ray image analyses: A machine vision based approach,'' 2016.

\bibitem{breiman2001random}
L.~Breiman, ``Random forests,'' {\em Machine learning}, vol.~45, no.~1,
  pp.~5--32, 2001.

\bibitem{brahim2019decision}
A.~Brahim, R.~Jennane, R.~Riad, T.~Janvier, L.~Khedher, H.~Toumi, and
  E.~Lespessailles, ``A decision support tool for early detection of knee
  osteoarthritis using x-ray imaging and machine learning: Data from the
  osteoarthritis initiative,'' {\em Computerized Medical Imaging and Graphics},
  vol.~73, pp.~11--18, 2019.

\bibitem{anderson1992explorations}
J.~R. Anderson and M.~Matessa, ``Explorations of an incremental, bayesian
  algorithm for categorization,'' {\em Machine Learning}, vol.~9, no.~4,
  pp.~275--308, 1992.

\bibitem{orlov2008wnd}
N.~Orlov, L.~Shamir, T.~Macura, J.~Johnston, D.~M. Eckley, and I.~G. Goldberg,
  ``Wnd-charm: Multi-purpose image classification using compound image
  transforms,'' {\em Pattern recognition letters}, vol.~29, no.~11,
  pp.~1684--1693, 2008.

\bibitem{shamir2008source}
L.~Shamir, N.~Orlov, D.~M. Eckley, T.~Macura, J.~Johnston, and I.~G. Goldberg,
  ``Source code for biology and medicine,'' {\em Source code for biology and
  medicine}, vol.~3, p.~13, 2008.

\bibitem{shamir2009early}
L.~Shamir, S.~M. Ling, W.~Scott, M.~Hochberg, L.~Ferrucci, and I.~G. Goldberg,
  ``Early detection of radiographic knee osteoarthritis using computer-aided
  analysis,'' {\em Osteoarthritis and Cartilage}, vol.~17, no.~10,
  pp.~1307--1312, 2009.

\bibitem{lee2010unsupervised}
H.~Lee, {\em Unsupervised feature learning via sparse hierarchical
  representations}, vol.~20.
\newblock Stanford University, 2010.

\bibitem{lecun1999object}
Y.~LeCun, P.~Haffner, L.~Bottou, and Y.~Bengio, ``Object recognition with
  gradient-based learning,'' in {\em Shape, contour and grouping in computer
  vision}, pp.~319--345, Springer, 1999.

\bibitem{sultana2018advancements}
F.~Sultana, A.~Sufian, and P.~Dutta, ``Advancements in image classification
  using convolutional neural network,'' in {\em 2018 Fourth International
  Conference on Research in Computational Intelligence and Communication
  Networks (ICRCICN)}, pp.~122--129, IEEE, 2018.

\bibitem{simonyan2014very}
K.~Simonyan and A.~Zisserman, ``Very deep convolutional networks for
  large-scale image recognition,'' {\em arXiv preprint arXiv:1409.1556}, 2014.

\bibitem{chatfield2014return}
K.~Chatfield, K.~Simonyan, A.~Vedaldi, and A.~Zisserman, ``Return of the devil
  in the details: Delving deep into convolutional nets,'' {\em arXiv preprint
  arXiv:1405.3531}, 2014.

\bibitem{jia2014caffe}
Y.~Jia, E.~Shelhamer, J.~Donahue, S.~Karayev, J.~Long, R.~Girshick,
  S.~Guadarrama, and T.~Darrell, ``Caffe: Convolutional architecture for fast
  feature embedding,'' in {\em Proceedings of the 22nd ACM international
  conference on Multimedia}, pp.~675--678, 2014.

\bibitem{chopra2005learning}
S.~Chopra, R.~Hadsell, and Y.~LeCun, ``Learning a similarity metric
  discriminatively, with application to face verification,'' in {\em 2005 IEEE
  Computer Society Conference on Computer Vision and Pattern Recognition
  (CVPR'05)}, vol.~1, pp.~539--546, IEEE, 2005.

\bibitem{dutta2019vgg}
A.~Dutta and A.~Zisserman, ``The vgg image annotator (via),'' {\em arXiv
  preprint arXiv:1904.10699}, 2019.

\bibitem{he2017mask}
K.~He, G.~Gkioxari, P.~Doll{\'a}r, and R.~Girshick, ``Mask r-cnn,'' in {\em
  Proceedings of the IEEE international conference on computer vision},
  pp.~2961--2969, 2017.

\bibitem{ren2015faster}
S.~Ren, K.~He, R.~Girshick, and J.~Sun, ``Faster r-cnn: Towards real-time
  object detection with region proposal networks,'' in {\em Advances in neural
  information processing systems}, pp.~91--99, 2015.

\bibitem{huang2017densely}
G.~Huang, Z.~Liu, L.~Van Der~Maaten, and K.~Q. Weinberger, ``Densely connected
  convolutional networks,'' in {\em Proceedings of the IEEE conference on
  computer vision and pattern recognition}, pp.~4700--4708, 2017.

\end{thebibliography}
\end{document}